\documentclass[10pt,letterpaper,conference]{IEEEtran}
\usepackage{maxim}
\usepackage{verbatim}
\mythmfalse

\def\cA{{\cal A}}
\def\cF{{\cal F}}
\def\cG{{\cal G}}
\def\cM{{\cal M}}
\def\cN{{\cal N}}
\def\cR{{\cal R}}
\def\cX{{\cal X}}

\begin{document}

\title{A Complexity-Regularized Quantization Approach to Nonlinear
  Dimensionality Reduction\vspace{-10pt}}
\author{\authorblockN{Maxim Raginsky}
\authorblockA{Beckman Institute and the University of Illinois\\
405 N Mathews Ave, Urbana, IL 61801, USA \\
Email: maxim@uiuc.edu\vspace{-10pt}}
}
\maketitle
\begin{abstract}We consider the problem of nonlinear dimensionality
reduction: given a training set of high-dimensional
data whose ``intrinsic'' low dimension is
assumed known, find a feature extraction map to low-dimensional space, a
reconstruction map back to high-dimensional space, and a geometric
description of the dimension-reduced data as a smooth manifold. We introduce a
complexity-regularized quantization approach for fitting a Gaussian
mixture model to the training set via a Lloyd algorithm. Complexity regularization controls the trade-off between
adaptation to the local shape of the underlying manifold and global geometric consistency. The
resulting mixture model is used to design the feature extraction
and reconstruction maps and to define a Riemannian metric on the
low-dimensional data. We also sketch a proof of
consistency of our scheme for the purposes of estimating the unknown
underlying pdf of high-dimensional data.\end{abstract}

\section{Introduction}
\label{sec:intro}

When dealing with high volumes of vector-valued data of some
large dimension $n$, it is often assumed that the data possess some intrinsic geometric description in a space of unknown dimension
$k < n$ and that the high dimensionality arises
from an unknown stochastic mapping of $\R^k$ into $\R^n$. We can pose
the problem of {\em nonlinear dimensionality reduction} (NLDR)
\cite{TenSilLan00, RowSau00}
as follows:
given raw data with values in $\R^n$, we wish to obtain optimal estimates of the intrinsic
dimension $k$ and of the stochastic map with the purpose of
modeling the intrinsic geometry of
the data in $\R^k$.

One typically considers the following set-up: we are given a sample $X^N
\equiv (X_1,\ldots,X_N)$, where $X_i$ are i.i.d. according to an unknown
absolutely continuous distribution $P^*$. The corresponding pdf $f^*$ has to be estimated from the observation as $\hat{f}_N \equiv
\hat{f}_N(X^N)$. The intrinsic dimension $k$ of the data may not be known
in advance and would also have be estimated as
$\hat{k}_N \equiv \hat{k}_N(X^N)$. Since the pdf $f^*$ is assumed to arise from a stochastic map of the low-dimensional space $\R^k$ into the high-dimensional space
$\R^n$, we can use our knowledge about $k$ and $f^*$ in order to make
inferences about the intrinsic geometry of the data. In the absence of
such knowledge, any such inference has to be
made based on the estimates $\hat{k}_N$ and $\hat{f}_N$. In this paper we introduce a complexity-regularized quantization
approach to NLDR, assuming that the intrinsic dimension $k$ of the data is given
(e.g., as a maximum-likelihood estimate \cite{LevBic05}). 

\section{Smooth manifolds and their noisy embeddings}
\label{sec:manifolds}

We begin with a quick sketch of some notions about smooth
manifolds \cite{Lan95}. A {\em smooth manifold} of dimension $k$ is a set $M$ together with a
collection $\cA = \{(U_l,\varphi_l) : l \in \Lambda\}$, where the sets
$U_l \subset M$ cover $M$ and each map $\varphi_l$ is a bijection of
$U_l$ onto an open set $\varphi_l(U_l)\subset\R^k$, such that for
all $l,l'$ with $U_l \cap U_{l'} \neq \varnothing$ the map
$\map{\varphi_{l'}\circ\varphi^{-1}_l}{\varphi_l(U_l \cap
  U_{l'})}{\varphi_{l'}(U_l \cap U_{l'})}$ is smooth. The pairs
$(U_l,\varphi_l)$ are called {\em charts} of $M$, and the entire
collection $\cA$ is referred to as an {\em atlas}. Intuitively, the
charts describe the points of $M$ by {\em local}
coordinates: given $p \in M$ and a chart $(U_l\ni p,\varphi_l)$,
$\varphi_l$ maps any point $q$ ``near $p$'' (i.e., $q \in U_l$) to an element of $\varphi_l(U_l) \subset \R^k$. Smoothness of the transition maps $\varphi_{l'}
\circ \varphi^{-1}_l$ ensures that local coordinates of a point transform
differentiably under a change of chart.

Assuming that $M$ is compact, we
can always choose the atlas $\cA$ in such a way that the indexing set
$\Lambda$ is finite and each $\varphi_l(U_l)$ is
an open ball of radius $r_l$ \cite[Thm.~3.3]{Lan95} (one can always
set $r_l \equiv 1$ for all $l \in \Lambda$, but we choose not to do
this for greater flexibility in modeling).

The next notion we need is that of a {\em tangent space} to $M$ at
point $p$, denoted by $T_pM$. Let $I \subset \R$ be an open interval
such that $0 \in I$. Consider the set of all curves
$\map{\xi}{I}{M}$ such that $\xi(0)=p$. Then for any chart $(U_l\ni p,\varphi_l)$ we have a
function $\map{\xi_l \deq \varphi_l \circ \xi}{I}{\R^k}$, such that
$\xi_l(t) \in \varphi_l(U_l)$ for all $t$ in a sufficiently small
neighborhood of $0$. We say that two such curves $\xi,\xi'$ are
equivalent iff $d\xi_{l,j}(t)/dt\big|_{t = 0} =
d\xi'_{l,j}(t)/dt\big|_{t=0}$, $j=1,\ldots,k$, for all $l \in \Lambda$ such that $U_l \ni p$, where $\xi_{l,j}(t)$
are the components of $\xi_l(t)$. The resulting set of equivalence classes
has the structure of a vector space of dimension $k$, and is precisely
the tangent space $T_pM$. Intuitively, $T_pM$ allows us to ``linearize'' $M$ around $p$. Note that, although all the
tangent spaces $T_pM,p\in M$ are isomorphic to each other and to
$\R^k$, there is no meaningful way to add elements of $T_pM$ and
$T_qM$ with $p,q$ distinct.

Next, we specify the class of stochastic embeddings dealt with in this
paper. Consider three random variables $L,Y,X$, where $L$ takes
values in the finite set $\Lambda$ with $w_l \deq \Pr(L=l)$, $Y$
takes values in $\R^k$, and $X$ takes values in $\R^n$. Conditional
distributions of $Y$ given $L$ and of $X$ given $Y,L$ are assumed to be
absolutely continuous and described by densities $f_{Y|L}$ and
$f_{X|YL}$, respectively. Since for a compact $M$ the images
$\varphi_l(U_l)$ of charts in $\cA$ are open balls of radii $r_l$, let
us suppose that the conditional mean $m_l(Y) \equiv \E[Y|L=l]$ is the center
of $\varphi_l(U_l)$ [we can therefore take $m_l(Y)=0$ for all $l \in
\Lambda$] and that the largest eigenvalue of the conditional covariance matrix $K_l(Y)
\equiv \E\big[YY^t\big| L=l\big]$ of $Y$ given $L=l$ is equal to $r^2_l$. It is convenient to think of the eigenvectors
$e^{(l)}_1,\ldots,e^{(l)}_k$ of $K_l(Y)$ as giving a basis of the
tangent space $T_{\varphi^{-1}_l(0)}M$. The unconditional density
$f_X$ of $X$ is the finite mixture
$f_X(x) = \sum_{l \in \Lambda}w_lf_l(x)$, where $f_l(x) \deq
\int_{\R^k}f_{X|YL}(x|y,l)f_{Y|L}(y|l)dy$. The resulting pdf follows the
local structure of the manifold $M$ and accounts both for
low- and high-dimensional noise.

As an example \cite{RowSauHin02}, let all $f_{Y|L}(y|l)$ be $k$-dimensional zero-mean Gaussians with unit covariance matrices,
$f_{Y|L}(y|l) = \cN(y;0,I) \equiv (2\pi)^{-k/2}\exp(-\frac{1}{2}y^ty)$, and
$f_{X|YL}(x|y,l) = \cN(x;\mu_l+A_ly,\Sigma_l)$, $\forall l \in
\Lambda$, for some means $\mu_l \in \R^n$, covariance matrices $\Sigma_l$, and
$n\times k$ matrices $A_l$, so that $f_X(x) = \sum_{l\in\Lambda}w_l\cN(x;\mu_l,A_lA^t_l + \Sigma_l)$.

\section{Complexity-regularized mixture models}
\label{sec:regularize}

Consider a random vector $X \in \R^n$ with an
absolutely continuous distribution $P_f$, described by a pdf $f$. We wish to find a
mixture model that would not only yield a good
``local'' approximation to $f$, but also have low complexity, where
the precise notion of complexity depends on application.

In order to set this up quantitatively, we use a complexity-regularized adaptation of the
quantizer mismatch approach of Gray and Linder \cite{GraLin03}. We
seek a finite
collection $\Gamma = \{g_m : m \in \cM \}$ of pdf's from a class $\cG$
of ``admissible'' models and a measurable
partition $\cR = \{R_m : m \in \cM\}$ of $\R^n$ that would minimize the
objective function
\begin{equation}
\bar{I}_f(\cR,\Gamma) \deq \sum_{m \in \cM}P_f(R_m)\big[ D(f_m\|g_m) +
  \mu \Phi_\Gamma(g_m)\big],
\label{eq:ibar}
\end{equation}
where $f_m$ is the pdf defined as $1_{\{x\in R_m\}}f(x)/P_f(R_m)$, $D(\cdot\|\cdot)$ is the relative
entropy, $\Phi_\Gamma(g_m)$ is a regularization functional that
quantifies the complexity of the $m$th model pdf relative to the entire
collection $\Gamma$, and $\mu \ge 0$ is the parameter that controls
the trade-off between the relative-entropy (mismatch) term and the complexity term.

This minimization problem can be posed as a {\em complexity-constrained quantization
  problem} with an encoder $\map{\alpha}{\R^n}{\cM}$ corresponding to
the partition $\cR = \{R_m\}$ through $\alpha(x) = m$ if $x \in R_m$, a decoder
$\map{\beta}{\cM}{\cG}$ defined by $\beta(m) = g_m$, and a length
function $\map{\ell}{\cM}{\{0,1,2,\ldots\}}$ satisfying the Kraft
inequality $\sum_{m \in \cM}e^{-\ell(m)} \le 1$. In order to
describe the encoder and to quantify the performance of the
quantization scheme, we need to choose a distortion measure between an input
vector and an encoder output in
such a way that minimizing average distortion would yield the
$\bar{I}$-functional (\ref{eq:ibar}) of the corresponding partition
and codebook.

Consider the
distortion $\rho(x,m) \deq \ln \big(f(x)/g_m(x)\big) + \ell(m) + \mu
\Phi_\Gamma(g_m)$ (this is not a distortion measure in the
strict sense since it can be negative, but its expectation with
respect to $f$ is nonnegative by the divergence inequality). For a given codebook $\Gamma$ and length function $\ell$, the optimal encoder is the
minimum-distortion encoder $\alpha(x) = \argmin_{m \in
  \cM}\rho(x,m)$ with ties broken arbitrarily. The resulting partition $\cR =
\{R_m\}$ yields the average distortion
\begin{eqnarray*}
&& \E_f \rho\big(X,\alpha(X)\big) = \sum_{m \in \cM}p_m\Big[\ell(m)+\mu
  \Phi_\Gamma(g_m) \\
&& \qquad \qquad + \int_{R_m}f_m(x)\ln\frac{p_mf_m(x)}{g_m(x)}dx\Big],
\end{eqnarray*}
where $p_m \deq P_f(R_m)$. Then
\begin{eqnarray*}
&& \E_f \rho\big(X,\alpha(X)\big)  = \sum_{m\in\cM}p_m\Big[D(f_m\|g_m)\\
&& \qquad \qquad +\ln \frac{p_m}{e^{-\ell(m)}} + \mu \Phi_\Gamma(g_m)\Big]\\
&& \qquad \ge  \sum_{m \in \cM}p_m\big[D(f_m\|g_m) + \mu\Phi_\Gamma(g_m)\big],
\end{eqnarray*}
with equality if and only if $\ell(m) =
-\ln p_m$. Thus, the optimal decoder and length function
for a given partition are such that the average $\rho$-distortion is precisely
the $\bar{I}$-functional. We can therefore iterate the
optimality properties of the encoder, decoder and length function in a
Lloyd-type descent algorithm; this can only decrease average distortion and thus
the $\bar{I}$-functional. Note that the $\ln f(x)$ term in $\rho(x,m)$ does not affect the
minimum-distortion encoder. Thus, as far as the encoder is concerned, the distortion measure $\rho_0(x,m) \deq -\ln g_m(x)
+ \ell(m) + \mu \Phi_\Gamma(g_m)$ is equivalent to $\rho$.

When the distribution of $X$ is unknown, we can take a sufficiently
large training sample $X^N =
(X_1,\ldots,X_N)$ and use a Lloyd
descent algorithm to empirically design a mixture model for the data:\smallskip

\noindent 1) {\bf Initialization:} begin with an initial codebook $\Gamma =
  \{g^{(0)}_m : m\in \cM\} \subset \cG$, where $\cG$ is the class of
  admissible models, and a length function
  $\map{\ell^{(0)}}{\cM}{\{0,1,2,\ldots\}}$. Set iteration number $r =
  1$, pick a convergence threshold $\epsilon$, 
  and let $D_0$ be the average $\rho_0$-distortion of the initial
  codebook.

\noindent 2) {\bf Minimum-distortion encoder:} encode each sample $X_i$
  into the index $\alpha^{(r)}(X_i) = \argmin_{m \in
    \cM}\rho_0(X_i,g^{(r-1)}_m)$.

\noindent 3) {\bf Centroid decoder:} update the codebook by
  minimizing over all $g \in \cG$ the empirical conditional expectation
$$
\E \big[ \rho_0(X,g) \big| \alpha^{(r)}(X)=m\big] \equiv \frac{1}{N^{(r)}_m} \sum_{i: \alpha^{(r)}(X_i) = m} \rho_0(X_i,g),
$$
where $N^{(r)}_m \deq \abs{\{ i : \alpha^{(r)}(X_i) = m\}}$, i.e., set $\beta^{(r)}(m) = g^{(r)}_m = \argmin_{g \in \cG}\E
\big[\rho_0(X,g) \big| \alpha^{(r)}(X)=m\big]$.

\noindent 4) {\bf Optimal length function:} if $N^{(r)}_m > 0$, let $\ell^{(r)}(m) = - \ln
  p^{(r)}_m$, where $p^{(r)}_m = N^{(r)}_m/N$. If $N^{(r)}_m = 0$, remove
  the corresponding cell from the code and decrease
  $\abs{\cM}$ by 1.

\noindent 5) {\bf Test:} compute the average $\rho$-distortion $D_r$ with
  the code $(\alpha^{(r)},\beta^{(r)},\ell^{(r)})$. If
  $(D_{r-1}-D_r)/D_{r-1} < \epsilon$, quit. Otherwise, go to Step 2 and
  continue.\smallskip

With a judicious choice of the initial codebook and length function,
this algorithm yields a finite mixture model $\{(g_m,p_m) :
m \in \cM\}$ as a good ``fit'' to the empirical distribution
of the data in the sense of near-optimal trade-off between the local
mismatch and complexity.

\section{Application to NLDR}
\label{sec:stochembed}

Given a training sample $X^N = (X_1,\ldots,X_N)$ of ``raw''
$n$-dimensional data and assuming its intrinsic dimension $k < n$ is
known, our goal is to
determine two mappings, $\map{v}{\R^n}{\R^k}$ and $\map{w}{\R^k}{\R^n}$,
where $v$ maps high-dimensional vectors to their dimension-reduced
versions and $w$ maps back to the high-dimensional space. In general,
the dimension-reducing map $v$ entails loss of information, so $w(v(x)) \neq x$. Therefore we will be interested
in the average distortion incurred by our scheme, $\bar{d}(v,w) \deq
\E[d(X,w(v(X)))]$, where $\map{d}{\R^n\times\R^n}{[0,\infty)}$ is a suitable distortion
measure on pairs of $n$-vectors, e.g., the squared Euclidean distance,
and the expectation is w.r.t. the empirical distribution of the
sample.

\subsection{Mixture model of a stochastic embedding}
\label{ssec:mmstochemb}

The first step is to use the above quantization scheme to fit a complexity-regularized
Gaussian mixture model to the training sample. Our class
$\cG$ of admissible model pdf's will be the set of all $n$-dimensional
Gaussians with nonsingular covariance matrices, $\cG = \{ \cN(x;\mu,K)
: \mu \in \R^n, \det K > 0 \}$, and for each finite set $\Gamma
\subset \cG$ we shall define a regularization functional
$\map{\Phi_\Gamma}{\Gamma}{[0,\infty)}$ that penalizes those $g \in
  \Gamma$ that are ``geometrically complex'' relative to the rest of
  $\Gamma$.

The idea of ``geometric complexity'' can be motivated
\cite{RowSauHin02,Bra03} by the example of the Gaussian
mixture model from Sect.~\ref{sec:manifolds}. The covariance matrix of the
$l$th component, $A_lA^t_l + \Sigma_l$, is invariant under
the mapping $A_l \mapsto A_lR$, where $R$ is a $k \times k$ orthogonal
matrix, i.e., $RR^t = I$. In geometric terms, a copy of the orthogonal
group $O_k$ associated with the
$l$th component of the mixture is the
group of rotations and reflections in the tangent space to $M$ at
$\varphi^{-1}_l(0)$. Thus, the log-likelihood term in $\rho_0$ is not affected by assigning arbitrary and
independent orientations to the tangent
spaces associated with the components of the mixture. However, since
our goal is to model the intrinsic {\em global}
geometry of the data, it should be possible to smoothly glue together the local data provided by our model. We therefore
require that the orientations of the tangent spaces at
``nearby'' points change smoothly as well. (In fact, one has to impose
certain continuity requirements on the orientation of the tangent
spaces in order to define measure and integration on the manifold
\cite[Ch.~XI]{Lan95}.)

Given a finite set $\Gamma \subset \cG$, we shall define the
regularization functional $\map{\Phi_\Gamma}{\Gamma}{[0,\infty)}$ as
\begin{equation}
\Phi_\Gamma(g) \deq \sum_{g' \in
  \Gamma\backslash\{g\}}\kappa(\mu_g,\mu_{g'})D(g'\|g),
\label{eq:gc}
\end{equation}
where $\map{\kappa}{\R^n\times\R^n}{\R^+}$ is a smooth positive
symmetric kernel such that $\kappa(x,x') \to 0$ as $\norm{x-x'} \to
\infty$, and
\begin{eqnarray*}
&& D(g'\|g) = \frac{1}{2}\big(\ln\det(K^{-1}_{g'}K_g) +
\tr(K^{-1}_gK_{g'}) \\
&& \qquad \qquad \qquad + (\mu_g-\mu_{g'})^tK^{-1}_g(\mu_g-\mu_{g'})-n\big)
\end{eqnarray*}
is the relative entropy between two Gaussians. Possible choices for
the kernel $\kappa$ are the inverse Euclidean distance $\kappa(x,x') =
\norm{x-x'}^{-1}$ \cite{Bra03a}, a Gaussian kernel $\kappa(x,x') =
\cN(x-x';0,\sigma^2I)$ for a suitable value of $\sigma$
\cite{Bra03,Bra03a} or a compactly supported ``bump'' $\kappa(x,x') =
\psi_{r_1,r_2}(x-x')$, where $\psi_{r_1,r_2}$ is an infinitely
differentiable reflection-symmetric function
that is identically zero everywhere outside a closed ball of radius
$r_2$ and one everywhere inside an open ball of radius $r_1 <
r_2$. The relative entropy serves as a measure of position and
orientation alignment of the tangent spaces, while the smoothing
kernel ensures that more weight is assigned to ``nearby''
components. This complexity functional is a generalization of the
``global coordination'' prior of Brand \cite{Bra03} to mixtures
with unequal component weights.

With these definitions of $\cG$ and $\Phi_\Gamma$, the
$\rho_0$-distortion for a codebook $\Gamma = \{g_m : m \in \cM\}$ and a
length function $\ell$ is
\begin{eqnarray*}
&& \rho_0(x,m) = \frac{1}{2}\ln\det K_m +
\frac{1}{2}(x-\mu_m)^tK^{-1}_m(x-\mu_m) \\
&& \quad \quad \quad + \ell(m) + \sum_{m' \in \cM\backslash\{m\}}
\kappa(\mu_m,\mu_{m'}) D(g_{m'}\|g_m),
\end{eqnarray*}
where we have also removed the $(n/2)\ln(2\pi)$ term as it does not
affect the encoder. The effect of the geometric complexity term is to
curve the boundaries of the partition cells according to
locally interpolated ``nonlocal information'' about the rest of the
codebook. Determining the Lloyd centroids for the decoder
will involve solving $\abs{\cM}$ simultaneous nonlinear equations for
the means and the same number of equations for the covariance
matrices. For computational efficiency we can use the kernel data from
the previous iteration, which would sacrifice optimality but avoid
nonlinear equations.

\subsection{Design of reduction and reconstruction maps}
\label{ssec:redrec}

The output of the previous step is a Gauss mixture model $\{ (g_m,p_m) : m \in
\cM\}$ and a partition $\cR = \{R_m\}$ of $\R^n$. Suppose that for
each $m \in \cM$ the eigenvectors
$e^{(m)}_1,\ldots,e^{(m)}_n$ of $K_m$ are
numbered in the order of decreasing eigenvalues, $\lambda^{(n)}_1 \ge
\ldots \ge \lambda^{(m)}_n$. The next step is to design the dimension-reducing map $v$ and the reconstruction
map $w$. One method, proposed by Brand \cite{Bra03}, is to use the
mixture model of the underlying pdf [obtained in his case by an EM algorithm with a prior
corresponding to the average of the complexity $\Phi_\Gamma(g)$ over the
entire codebook and with equiprobable components of the mixture] to
construct a mixture of local affine transforms, preceded by
local Karhunen-\Loeve\ transforms, as a solution to a weighted
least-squares problem.

However, we can use the encoder partition $\cR$ directly: for each $m
\in \cM$, let $v_m(x) \deq
\Pi_m(x-\mu_m)$, where $\Pi_m$ is the projection onto the first $k$
eigenvectors of $K_m$, and then define $v(x) = \sum_{m \in \cM}1_{\{x\in
  R_m\}}v_m(x)$. This approach is similar to local principal component
analysis of Kambhatla and Leen \cite{KamLee97}, except that their
quantizer was not complexity-regularized and therefore the shape of
the resulting Voronoi regions was determined only by local statistical
data. We can describe the operation of dimension reduction (feature
extraction) as an encoder $\map{\hat{v}}{\R^n}{\cM\times\R^k}$, so that $\hat{v}(x) =
(\alpha(x),v_{\alpha(x)}(x))$, where $\alpha$ is the
minimum-distortion encoder for the $\rho_0$-distortion.

The corresponding reconstruction operation can be
designed as a decoder $\map{\hat{w}}{\cM \times \R^k}{\R^n}$ which
receives a pair $(m,u)$, $m \in \cM,u \in \R^k$, and computes $w_m(u) =
\mu_m + \sum^k_{i=1}\ave{u,e^{(m)}_i}e^{(m)}_i$, where
$\ave{\cdot,\cdot}$ denotes the usual scalar product in $\R^k$.

This encoder-decoder pair is a composite Karhunen-\Loeve\ transform coder matched to the mixture source $g = \sum_m
p_mg_m$. If the data alphabet $\cX$ is compact, then the squared-error
distortion is bounded by some $A > 0$,
and the mismatch due to using this composite coder on the disjoint
mixture
source $f = \sum_m p_mf_m$ can be bounded from above by $A\|f-g\|_1$, where
$\|\cdot\|_1$ is the $L_1$ norm. Provided that the mixture $g$ is
optimal for $f$ in the sense of minimizing the $\rho$-distortion,
we can use Pinsker's inequality
\cite[Ch.~5]{DevLug01} $\|f-g\|_1 \le \sqrt{2D(f\|g)}$ and convexity of
the relative entropy to further bound the mismatch by $A\sqrt{2\big(\bar{I}_f(\cR,\Gamma) - \mu\sum_mp_m\Phi_\Gamma(g_m)\big)}$.

Note that the maps
$v$ and $w$ are not smooth, unlike the analogous maps of Brand \cite{Bra03,Bra03a}. This
is an artifact of the hard partitioning used in our scheme. However,
hard partitioning has certain advantages: it allows
for use of composite codes \cite{GraLin03} and nonlinear interpolative vector quantization
\cite{Ger90} if additional compression of dimension-reduced data is
required. Moreover, the lack of smoothness is not a problem in our
case because
we can use kernel interpolation techniques to model the geometry of
dimension-reduced data by a smooth manifold, as explained next.

\subsection{Manifold structure of dimension-reduced data}
\label{ssec:dimredman}

Our use of mixture models has been motivated by
certain assumptions about the structure of stochastic embeddings of
low-dimensional manifolds into high-dimensional spaces. In
particular, given an $n$-dimensional Gaussian mixture model $\{(g_m,p_m) : m \in \cM\}$, we can associate to each component of the
mixture a chart of the underlying manifold, such that the image of the
chart in $\R^k$ is an open ball of radius $r_m = (\lambda^{(m)}_1)^{1/2}$ centered at the origin, and we can take the
first $k$ eigenvectors of the covariance matrix of $g_m$ as coordinate
axes in the tangent space to the manifold at
the inverse image of $0 \in \R^k$ under the $m$th chart. Owing to
geometric complexity regularization, the orientations of tangent
spaces change smoothly as a function of position.

Ideally, one would like to construct a smooth manifold consistent with the
given descriptions of charts and tangent spaces. However, this is a
fairly difficult task since we not only have to define a smooth coordinate map
$\varphi_m$ for each chart, but also make sure that these maps satisfy
the chart compatibility condition. Instead, we can construct the
manifold {\em implicitly} by gluing the coordinate frames of the
tangent spaces into an object having a smooth inner product.

Specifically, let us fix a sufficiently small $\delta > 0$, and let
$\psi_m$ be an infinitely differentiable function
that is identically zero everywhere outside a closed ball of radius
$r_m$ and one everywhere inside an open ball of radius $r_m-\delta$, with
both balls centered at $\Pi_m\mu_m$. Let $\eta_m(u) \deq \frac{p_m\psi_m(u)}{\sum_{m \in
    \cM}p_m\psi_m(u)}$. The inner product of two vectors $u,u' \in
\R^k$, treated as elements of the tangent space
$T_{\varphi^{-1}_m(0)}M$, is given by $\ave{u,u'}_m = \sum^k_{i=1} \ave{u,e^{(m)}_i}\ave{e^{(m)}_i,u'}$. Then for each $y \in \R^k$ the map $\map{g_y}{\R^k \times \R^k}{[0,\infty)}$,
$$
g_y(u,u') \deq \sum_{m \in \cM}\eta_m(y+\Pi_m\mu_m)\ave{u,u'}_m,
$$
is a symmetric form, which is positive definite
whenever $\eta_m(y+\Pi_m\mu_m) \neq 0$ for at least one value of
$m$. In addition, the map $y \mapsto g_y(\cdot,\cdot)$ is smooth. In
this way, we have implicitly defined a {\em Riemannian metric}
\cite[Ch.~VII]{Lan95} on the underlying
manifold. The functions $\eta_m$ form a so-called {\em smooth
  partition of unity}, which is the only known
way of gluing together local geometric data to form smooth objects \cite[Ch.~II]{Lan95}.

In geometric terms, $\eta_m(y+\Pi_m\mu_m) = 0$ for all $m$ if and only
if $y \in \R^k$ is an image under the dimension-reduction map of a
point in $\R^n$ whose first $k$ principal components w.r.t. each
Gaussian in the mixture model fall outside the covariance ellipsoid of
that Gaussian. If the mixture model is close to
optimum, this will happen with negligible probability. A practical
advantage of this feature of our scheme is in rendering it robust to outliers.

\section{Consistency and codebook design}
\label{sec:consistency}

Our mixture modeling
scheme can also be used to estimate the ``true'' but unknown pdf $f^*$ of the
high-dimensional data, if we assume that $f^*$ belongs to some fixed
class $\cF$. Indeed, the empirically designed codebook
$\Gamma = \{g_m : m \in \cM\}$ of Gaussian pdf's, the corresponding
component weights $\{p_m\}$, and the mixture $g = \sum_{m \in
  \cM}p_mg_m$ are random variables since they depend on the training sample $X^N$. We are interested in the quality of
approximation of $f^*$ by the mixture $g \equiv g(X^N)$.

Following Moulin and Liu \cite{MouLiu00}, we use the relative-entropy loss function
$D(f^*\|g)$. We shall give an upper bound on the loss in terms of the {\em index of
  resolvability} \cite{MouLiu00}
$$
R_{\mu,N}(f^*) \deq \Min_{m \in \cM}\left[D(f^*\|g_m)+\frac{\mu
    L(g_m)}{N}\right],
$$
where $L(g_m) \deq \Phi_\Gamma(g_m) - \ln
p_m$, which quantifies how well $f^*$ can be
approximated, in the relative-entropy sense (and, by Pinsker's
inequality, in $L_1$ sense),
by a Gaussian of moderate geometric complexity relative to the rest of
the codebook. We have the following result:

\begin{theorem} Let the codebook $\Gamma = \{g_m : m \in \cM\}$ of
  Gaussian pdf's be such that the log-likelihood ratios $U_m \deq -\ln
  \big(f^*(X)/g_m(X)\big)$ uniformly satisfy the {\em Bernstein moment condition} \cite{DevLug01}, i.e.,
  there exists some $h > 0$ such that $\E\abs{U_m-\E U_m}^k \le
  (1/2)\var(U_m)k!h^{k-2}$ for all $k \ge 2$. Let $M(f^*)$ be the
  smallest number such that $\var(U_m) \le -M(f^*)\E U_m$ for all $m
  \in \cM$
(owing to the Bernstein condition, it is nonnegative and
  finite). Then, for any $\mu > h + M(f^*)/2$ and $\delta > 0$, 
\begin{equation}
\Pr\left\{D(f^*\|g) \le \frac{1+\alpha}{1-\alpha}R_{\mu,N}(f^*) + \frac{2\mu
  \ln\frac{\abs{\cM}}{\delta}}{(1-\alpha)N}\right\} \ge 1-2\delta,
\label{eq:lossbound1}
\end{equation}
where $\alpha = \frac{M(f^*)}{2(\mu - h)}$. The expected loss satisfies
\begin{equation}
\E[D(f^*\|g)] \le \frac{1+\alpha}{1-\alpha}R_{\mu,N}(f^*) +
\frac{4\abs{\cM}\mu}{(1-\alpha)N}.
\label{eq:lossbound2}
\end{equation}
The probabilities and expectations are all w.r.t. the pdf $f^*$.\end{theorem}

\begin{proof} Due to the fact that $\Phi_\Gamma(g_m) \ge 0$ for all $m
  \in \cM$, the composite complexity $L(g_m)$ satisfies the Kraft inequality. Then we
can use a strategy similar to that of Moulin and Liu \cite{MouLiu00} to prove that
$$
\Pr \left\{D(f^*\|g_m) \ge \frac{1+\alpha}{1-\alpha}R_{\mu,N}(f^*) +
\frac{2\mu\ln\frac{\abs{\cM}}{\delta}}{(1-\alpha)N}\right\} \le
\frac{2\delta}{\abs{\cM}}
$$
for each $m \in \cM$. Hence, by the union bound
$$
D(f^*\|g_m) \le \frac{1+\alpha}{1-\alpha}R_{\mu,N}(f^*) +
\frac{2\mu\ln\frac{\abs{\cM}}{\delta}}{(1-\alpha)N}
$$
for all $m \in \cM$, except for an event of probability at most
$2\delta$. By convexity of the relative entropy,
$D(f^*\|g_m) \le C$ for all $m \in \cM$ implies that $D(f^*\|g) \le C$
for $g = \sum_{m \in \cM}p_mg_m$. Therefore
$$
D(f^*\|g) \le \frac{1+\alpha}{1-\alpha}R_{\mu,N}(f^*) +
\frac{2\mu \ln \frac{\abs{\cM}}{\delta}}{(1-\alpha)N}
$$
with probability at least $1-2\delta$. To prove (\ref{eq:lossbound1}), we use the fact \cite{DevLug01} that
if $Z$ is a random variable with $\E\abs{Z} < \infty$, then $\E[Z] \le
\int^\infty_0 \Pr[Z\ge t]dt$. We let $Z =
D(f^*\|g)-\frac{1+\alpha}{1-\alpha}R_{\mu,N}(f^*)$ and choose $\delta
= \abs{\cM}e^{-\frac{Nt(1-\alpha)}{2\mu}}$. Then $\E[Z] \le \frac{4\abs{\cM}\mu}{(1-\alpha)N}$, which proves (\ref{eq:lossbound2}).
\end{proof}

To discuss consistency in the large-sample limit, consider a
sequence of empirically designed mixture models
$\{(g^{(N)}_m,p^{(N)}_m) : m \in \cM^{(N)}\}$. This is
different from the usual empirical quantizer design, where we
increase the training set size but keep the number of quantizer
levels fixed. The scheme is consistent
in the relative-entropy sense if $\E D(f^*\|g^{(N)}) \to 0$ as $N \to
\infty$, where $g^{(N)} = \sum_{m\in\cM^{(N)}}p^{(N)}_mg^{(N)}_m$ and the
expectation is with respect to $f^*$.

A sufficient condition for consistency can be determined by inspection
of the upper bound in Eq.~(\ref{eq:lossbound2}). Specifically, we
require that the codebooks $\Gamma^{(N)}$ satisfy: (a) $\max_{m \in
  \cM^{(N)}}L(g^{(N)}_m) = o(N)$, (b) $\min_{m \in
  \cM^{(N)}}D(f^*\|g_m) = o(1)$ for all $f^* \in \cF$, and (c)
$\abs{\cM^{(N)}} = o(N)$. Condition (c) can be satisfied by
initializing the Lloyd algorithm by a codebook of size much smaller than the training set size $N$, which is usually done in practice in order
to ensure good training performance. The first two conditions can also
be easily met in many practical settings.

Consider, for instance, the class $\cF$ of all pdf's supported on a
compact $\cX \subset \R^n$ and Lipschitz-continuous with Lipschitz constant $c$. Then, if we take as our class of
admissible Gaussians $\cG = \{\cN(x;\mu,K) : \mu \in \cX, c_1 \le \det K
\le c_2\}$ for suitably chosen constants $c_1,c_2 > 0$
independent of $N$, the relative entropy $D(g\|g')$ of any two $g,g'
\in \cG$ can be bounded independently of $N$, and
condition (a) will be met with proper choice of the component
weights. Condition (b) is likewise easy to meet since the maximum
value of any $f^* \in \cF$ depends only on the set $\cX$, the
Lipschitz constant $c$, and the dimension $n$.

In general, the issue of optimal codebook design is closely related to
the problem of universal vector quantization \cite{ChoEffGra96}:
we can consider, e.g., a class $\cF$ of pdf's with disjoint
supports contained in a compact $\cX \subset \R^n$. Then a sequence of
Gaussian codebooks that yields a consistent estimate of each $f^* \in \cF$
in the large-sample limit is weakly minimax universal
\cite{ChoEffGra96} for
$\cF$ and can also be used to quantize any source contained in the
$L_1$-closed convex hull of $\cF$.

\section{Discussion}
\label{sec:discuss}

We have introduced a complexity-regularized quantization approach to
NLDR. One advantage of this scheme over existing methods for NLDR
based on Gaussian mixtures, e.g., \cite{Bra03}, is that, instead of
fitting a Gauss mixture to the entire sample, we design a codebook
of Gaussians that provides a good trade-off between local adaptation to
the data and global geometric coherence, which is key to robust
geometric modeling. Complexity regularization is based on a kernel
smoothing technique that allows for a meaningful geometric
description of dimension-reduced data by means of a Riemannian metric
and is also robust to outliers. Moreover, to our knowledge, the
consistency proof presented here is the first theoretical asymptotic
consistency result applied to NLDR.

Work is currently underway to implement the proposed scheme for applications to image processing and computer vision. Also
planned is future work on a quantization-based approach to
estimating the intrinsic dimension of the data and on assessing asymptotic
{\em geometric} consistency of our scheme in terms of the
Gromov-Hausdorff distance between compact metric spaces
\cite{Pet90}.\smallskip

\noindent{\bf Acknowledgment.} I would like to thank Svetlana Lazebnik and Prof. Pierre Moulin for useful discussions. This research has
been supported by the Beckman Postdoctoral Fellowship.

\end{document}